\documentclass[a4paper,fleqn]{cas-dc}

\usepackage{multirow}
\usepackage{cleveref}
\usepackage{amssymb}
\usepackage{placeins}
\usepackage{pifont}
\usepackage{url}
\usepackage{flushend}
\usepackage[numbers]{natbib}
\usepackage[export]{adjustbox}

\hypersetup{draft}
\begin{document}
\let\WriteBookmarks\relax
\def\floatpagepagefraction{1}
\def\textpagefraction{.001}

\shorttitle{Detecting Vocal Fatigue with Neural Embeddings}

\shortauthors{S. P. Bayerl, D. Wagner et~al.}

\title[mode = title]{Detecting Vocal Fatigue with Neural Embeddings}

\author[1]{Sebastian P. Bayerl}[orcid=0000-0002-3502-9511]
\fnmark[1]
\cormark[1]
\author[1]{Dominik Wagner}
\fnmark[1]
\cormark[1]
\author[1]{Ilja Baumann}
\cormark[1]
\author[1]{Tobias Bocklet}
\cormark[1]
\author[1]{Korbinian Riedhammer}[orcid=0000-0003-3582-2154]
\cormark[1]

\address[1]{Technische Hochschule N\"urnberg Georg Simon Ohm, Germany, Keßlerplatz 12, Nuremberg, 90489, Germany}


\cortext[cor1]{E-mail: \{firstname\}.\{lastname\}@th-nuernberg.de}

\fntext[fn1,fn2]{Equal contribution, listed in alphabetical order.}

\begin{abstract}
Vocal fatigue refers to the feeling of tiredness and weakness of voice due to extended utilization. 
This paper investigates the effectiveness of neural embeddings  for the detection of vocal fatigue. 
We compare x-vectors, ECAPA-TDNN, and wav2vec 2.0 embeddings on a corpus of academic spoken English.
Low-dimensional mappings of the data reveal that neural embeddings capture information about the change in vocal characteristics of a speaker during prolonged voice usage.  
We show that vocal fatigue can be reliably predicted using all three types of neural embeddings after 40 minutes of continuous speaking when temporal smoothing and normalization are applied to the extracted embeddings. 
We employ support vector machines for classification and achieve accuracy scores of 81\% using x-vectors, 85\% using ECAPA-TDNN embeddings, and 82\% using wav2vec 2.0 embeddings as input features. 
We obtain an accuracy score of 76\%, when the trained system is applied to a different speaker and recording environment without any adaptation. 

\end{abstract}


\maketitle

\section{Introduction}
Vocal fatigue is a common phenomenon among teachers \citep{Gotaas1993teacher}, singers and actors \citep{Benninger2010}, as well as call center agents \citep{LEHTO2008callcenter}. 
Even though, there is no universally accepted definition, vocal fatigue has been commonly described as the feeling of tiredness and weakness of voice due to extended utilization of the vocal apparatus \citep{Welham2003,Nanjundeswaran2015}. 
It can be distinguished by the type of diagnosis method, where ``performance fatigue'' refers to a measurable limitation in vocal performance, and ``perceived fatigue'' refers to a subjective assessment of the impact that limited vocal performance has on the patient's daily activities \citep{Hunter2020-qx}. 
Currently, vocal fatigue is often diagnosed based on subjective features in the vocal fatigue index (VFI), a standardized questionnaire consisting of 19 questions with the aim of differentiating between individuals who suffer from vocal fatigue and those who do not \citep{Nanjundeswaran2015}. 
Our work is a contribution to the objective evaluation of ``performance fatigue''. 
The ability to accurately predict the occurrence of this aspect of vocal fatigue could potentially support voice professionals such as teachers or singers in their work and help to avoid over-utilization of their voice by monitoring the current state of fatigue. 

Following Caraty and Montacié \cite{CARATY2014453}, who employ a similar approach, we assume that vocal fatigue can be measured by observing the change of voice characteristics over time. 
In this work, we utilize three state-of-the-art systems to extract latent speech representations to encode voice characteristics and explore their suitability for the detection of vocal fatigue. 
We employ a pre-trained wav2vec 2.0 (W2V2) \citep{baevski_wav2vec_2020} encoder, an x-vector system \citep{snyder_x-vectors_2018}, and an ECAPA-TDNN \citep{Desplanques2020ecapa} for feature extraction. 
We visualize the extracted neural embeddings by mapping them into two-dimensional space using t-distributed stochastic neighbor embedding (t-SNE) \citep{vandermaaten08tsne}.
We demonstrate that ``performance fatigue'' can be predicted by using support vector machine (SVM) \citep{boser92svm} classifiers and that recording-level normalization, as well as temporal smoothing, can significantly improve classification performance. 

The main goal of this paper is the automatic detection of ``performance fatigue'' in the voice of a professional academic speaker after prolonged voice utilization giving lectures based on characteristics associated with the speaking time that are captured via high-dimensional latent representations.

\section{Related Work}
This work employs learned representations as features for the automatic detection of ``performance fatigue''. 
These learned representations are generated from audio features. 
Previous studies relating various types of features obtained from audio signals to vocal fatigue are discussed in \Cref{ssec:work_fatigue}. 
Studies on vocal fatigue detection with machine learning approaches are summarized in \Cref{ssec:work_detect}, and methods to generate latent speech representations, as well as their relation to conventional audio features is discussed in \Cref{ssec:work_representation}.
\subsection{Relationship Between Vocal Fatigue and Audio Features}\label{ssec:work_fatigue}
Previous studies attempted to measure changes in prosodic features, such as estimates of fundamental frequency ($F0$) or sound pressure level over longer time periods and applied statistical tests to determine the significance of those changes \citep{CARATY2014453, Laukkanen2008, REMACLE2012e177}.
Most works found that an increase in $F0$ and voice intensity level is correlated with vocal fatigue through extended voice usage. 
Additionally, prosodic feature changes were found to be correlated with questionnaire-based self-assessments of vocal fatigue \citep{Solomon2003,Laukkanen2008}. 
The work by Laukkanen et al. \cite{Laukkanen2008} investigated the relationship between reported symptoms of vocal fatigue (``perceived fatigue'') and measurable acoustic variables (``performance fatigue''). 
The subjects, who were experienced vocal professionals, completed a questionnaire at the beginning and the end of their working day. 
The authors found elevated measurements for $F0$ and sound pressure level after working days, on which the subjects preceived more tiredness of throat.
However, the authors concluded that the differences in acoustic parameters mainly reflect the increased muscle activity as a result of vocal loading. 
Their questionnaire used voice quality, difficulty of phonation and tiredness of throat as symptoms of vocal fatigue. 

A similar study by Remacle et al. \cite{REMACLE2012e177} observed significant differences in $F0$ measured at the beginning of a prolonged reading session and later measurements in non-professional voice users. 
Their results indicated that vocal features such as an elevated average $F0$ coincides with increased self-reported vocal fatigue ratings, as well as more laryngeal discomfort during periods of prolonged reading. 
Vocal fatigue was defined as ``tiredness of voice or in neck muscles'' in their questionnaire. 

Carroll et al. \cite{Carroll2006-xd} obtained prosodic features such as $F0$, voicing time, and sound pressure level, as well as subjective ratings in 2-hour intervals from professional singers via a Pocket PC attached to the subjects' bodies (Dosimeter). 
Their questionnaire included voice quality and vocal effort as indicators for vocal fatigue. 
The authors found that the subjects' inability to produce soft voice and vocal effort as measured by the Dosimeter often increased after practice sessions. 

A similar study by Lei et al. \cite{Lei2020-hl} showed that the overall severity of vocal fatigue measured by a combination of 15 prosodic features strongly increased after two consecutive 25-minute sessions of loud reading.
The primary subjective indicator for vocal fatigue in this study was voice quality, evaluated with the Consensus Auditory-Perceptual Evaluation of Voice (CAPE-V) and the Self-Administrated Voice Rating (SAVRa) by the American Speech-Language and Hearing Association. 

The studies \cite{CARATY2014453} and \cite{Caraty2010MultivariateAO} do not rely on subjective fatigue ratings but analyze acoustic features in the context of vocal fatigue. 
In \citep{CARATY2014453}, the performance of a prosodic analysis and a phoneme-based analysis is compared to a binary SVM trained on a large audio feature set.  
The study showed that a simple prosodic analysis is often not enough (e.g., increased levels of $F0$ were found in only one of their four test subjects), but using a classifier trained on a comprehensive set of audio features, as well as the phoneme-based analysis yielded promising results. 

The authors of \cite{Caraty2010MultivariateAO} performed an analysis of spectral and prosodic changes in phonemes over prolonged voice usage. 
They conclude that spectral parameters are more predictive of vocal fatigue than prosodic parameters. 
The results were verified with an analysis of reading errors and speech dysfluencies. 
A significant increase in the disfluency rate (e.g., word substitutions and repetitions) during long periods of reading was also observed.

In experimental setups, vocal fatigue is usually triggered by a continuous reading task. 
However, parameters such as task duration, reading content, environment and the number of readers varies depending on the study. 

\subsection{Automatic Vocal Fatigue Detection}\label{ssec:work_detect}
The automatic detection of vocal fatigue in prolonged human speech has been the subject of study in multiple previous works.
Most works employ hand-crafted features, such as spectrograms, cepstral coefficients, and fundamental frequency, that are directly obtained from the audio signal.  

In addition to their statistical analysis, Caraty and Montacié used SVM classifiers on 1582 prosodic features \cite{CARATY2014453} . 
Their data comprised three hour long recordings of read speech, which were split into an early segment (first 30 minutes) and a late segment (last 30 minutes) to discriminate between the classes ``fatigue'' and ``non-fatigue''.  
In \cite{SHEN2021403}, latent features extracted from an autoencoder model, are used to to classify speech into ``fatigue'' and ``non-fatigue'' with SVMs.  
The work by Gao et al. \cite{gao2021semg} did not rely on acoustic features;
Instead, sensor data from surface electromyography was used to detect vocal fatigue with SVMs.

\subsection{Latent Representations}\label{ssec:work_representation}
Another approach for feature extraction that has shown great promise in recent years, utilizes internal representations of neural networks.
These latent features have been successfully applied in automatic speech recognition (ASR) \citep{baevski_unsupervised_2021} and speech classification tasks, such as speaker and language identification \citep{snyder_x-vectors_2018, snyder18_odyssey,tjandra2021improved,fan21_interspeech}, phoneme recognition, speech emotion recognition, and dysfluency detection \citep{baevski_wav2vec_2020,pepino_emotion_2021,bayerl_DetectingDysfluenciesStuttering_2022a}. 

One of the first approaches to generate speaker representations from audio signals was a statistical model that uses factor analysis as a feature extractor \citep{dehak2011ivector}. 
Each recording was represented by a compact vector called i-vector. 
The i-vector system has been superseded by internal representations of neural networks trained on speaker- or language classification tasks. 
Widely used neural representations are x-vectors \citep{snyder_x-vectors_2018} and ECAPA-TDNN embeddings \citep{Desplanques2020ecapa}. 
Recently, more generic and universally applicable embeddings obtained from wav2vec 2.0 (W2V2) models \citep{baevski_wav2vec_2020} have been used. 

However, the extent to which neural embeddings can capture the changes in a speaker's voice during prolonged usage remains to be explored. 
To the best of our knowledge, there have been no previous attempts to leverage latent neural features to visualize and detect vocal fatigue. 

In \cite{pmlr-v139-weston21a},  latent speech representations obtained from models such as wav2vec 2.0 and x-vector have been shown to be well-suited to capture non-timbral prosody, i.e., the remaining components after removing speaker characteristics. 
Parts of prosody such as pitch, rhythm and tempo are reflected in $F0$, energy or intensity, and the speech rate respectively. 
The same components have been extensively used as features in previous works to monitor and detect vocal fatigue \citep{CARATY2014453,Caraty2010MultivariateAO,Xue2019-rb,Dhaeseleer2016-rl}. 
Therefore, we hypothesize that neural speech representations are capable of capturing characteristics required to detect vocal fatigue.

\section{Data}\label{sec:data}
We use the audio part of the LMELectures multimedia corpus of academic spoken English by Riedhammer et al. \cite{riedhammer13lme} to conduct our experiments. 
The corpus consists of recordings from 36 lectures covering pattern analysis, machine learning, and medical image processing. 
The main distinction of the LMELectures to other corpora of academic spoken English is its constant recording environment, the single speaker, and the narrow range of topics. 

The corpus contains recordings from two distinct graduate-level courses titled pattern analysis (\textit{PA}) and interventional medical image processing (\textit{IMIP}). 
The lectures were read in the same year by a non-native but proficient male speaker. 
All recordings were acquired in the same room using the same close-talking microphone. 
The microphone reduced a large portion of the room's acoustics and background noises. 
Nevertheless, the recordings were professionally edited afterward, to ensure a constant high audio quality throughout all lectures. 
The LMELectures corpus is well suited to measure vocal fatigue, since the recordings contain sufficiently long uninterrupted spontaneous speech by a single person in high quality. 
Furthermore, giving lectures as a non-native speaker is an activity that leads to a high cognitive load; a factor that has also been shown to contribute to vocal fatigue. 

We exclude all lectures shorter than 60 minutes.
The remaining 19 lectures (10 \textit{IMIP} and 9 \textit{PA}) amount to 27 hours of audio material. 
The duration of this subset varies between 67 and 91 minutes, with a mean of 84 minutes. 
The remaining 19 lectures were recorded in the morning on different days.
The lecturer had no prior lectures on these days. 

We create an additional corpus to test the generalization ability of our approach. 
The additional test corpus consists of recordings from a graduate-level course on deep learning (DL), which was read in English by a non-native but proficient male speaker, who has roughly the same age as the speaker in the LMELectures.\footnote{Deep Learning lectures by Andreas Maier are licensed under \mbox{CC BY} 4.0, available at \protect \url{https://www.fau.tv/course/id/662}}
The lectures were recorded in a different  but constant recording environment.  
The DL corpus consists of 12 lectures, from which 2 lectures were removed due to their duration of less than 60 minutes.
The total duration of the remaining 10 recordings is 12.5 hours, with a mean of 75 minutes. 
We prefer the creation of the DL lecture corpus over other available corpora of academic spoken English because of its similarity to the LMELectures in terms of lecture length and recording conditions. 
\subsection{Limitations}\label{ssec:limitations}
It is important to point out that the data used in our experiments has limitations that may reduce the overall generalizability of the approach. 
The main corpus used for training and testing contains a single voice by a non-native male speaker. 
Furthermore, the recordings cover a narrow range of topics and the recording environment is constant. 
While these features are beneficial to ensure that the classification results are not driven by factors other than the change in the speaker's vocal characteristics, such as changes in the recording environment or speaking style, the generalizability across a wide range of speakers and conditions cannot be ensured. 
The additional corpus used for robustness testing is also similar to the main corpus in several ways (male, similar age, similar lecture topics) and therefore provides only limited insights into the overall applicability. 
Therefore, we cannot make statements about the influence of changes in recording environment (e.g. different microphone and room acoustics), speaking style (e.g. conversational speech instead of read lectures), as well as demographic factors such as gender, age, and accent of the speakers. 

\section{Method}
Our approach relies on the extraction of latent representions from neural networks that have been optimized for speaker classification and automatic speech recognition. 
We introduce the general idea behind this approach and describe the systems used for feature extraction in \Cref{sec:representation}. 
Our baseline method for obtaining prosodic features is introduced in \Cref{sec:opensmile}. 
Finally, two methods of data transformation and aggregation, which we found crucial in improving the performance of a our classifiers are described in \Cref{ssec:smooth}. 
Unless otherwise stated, we use the terms ``performance fatigue'' and vocal fatigue interchangeably for the remainder of this paper.
\subsection{Representation Learning}\label{sec:representation}
The performance of machine learning algorithms depends on the way input data is represented.
Different feature representations can entangle and hide different explanatory factors of variation in the data \citep{bengio13representation}. 
Traditionally, domain-specific knowledge has been leveraged to design useful representations for specific tasks. 
However, this approach is costly and requires manual effort. 
Therefore, more generic methods to extract discriminative information from the data without the need for extensive prior knowledge, have been developed in recent years. 

Representation learning is the task of learning features from the input data that allows classifiers or other supervised predictors to extract useful information. 
Ideally, the representation captures the posterior distribution of the underlying explanatory components for the observed input data \citep{bengio13representation}.

In modern systems, latent representations (embeddings) are obtained at one of the hidden layers of the underlying neural network.
Their optimization criterion varies depending on the intended usage of the learnt representations. 
A system intended for language identification has the training objective to discriminate between different languages, whereas a system intended for speaker identification learns to discriminate between utterances from different speakers. 
The classification is usually accomplished by computing similarity scores between pairs of embeddings; one embedding obtained from the input recording and multiple prototype embeddings representing the different classes available. 

In this work, we compare three state-of-the-art systems to extract neural embeddings: x-vector, ECAPA-TDNN, and wav2vec 2.0.

\subsubsection{x-vector}\label{sec:xvec}
The x-vector architecture \citep{snyder_x-vectors_2018} is a \textit{time delay neural network} (TDNN) that aggregates variable-length inputs across time to create fixed-length representations capable of capturing speaker characteristics. 
Speaker embeddings are extracted from a bottleneck layer prior to the output layer. 

We follow the data preparation steps provided by the \texttt{voxceleb/v2} recipe of the Kaldi toolkit \citep{povey11}. 
However, our model slightly deviates from the architecture in the recipe since we removed the frame limit in the statistics pooling layer. 
This enables us to apply the pooling operation on an entire lecture without  computing the average over multiple parts of the recording.  
We used the x-vectors generated over the entire length of each lecture to remove global traits from other x-vectors that cover shorter subsequences of the lecture.
We do this to enhance the characteristics related to the speaker's voice. 
This approach is discussed in more detail in \Cref{ssec:smooth}. 

The model is trained on the VoxCeleb \citep{Nagrani17} dataset, which contains approximately 1.2 million utterances from 7,323 different speakers. 
The training data is augmented with additive noises from the MUSAN corpus \citep{musan2015} and reverberation using a collection of room impulse responses \citep{rirs2017}. 
The input features are 30-dimensional MFCCs using a frame width of 25 ms and a frame-shift of 10 ms. 
\subsubsection{ECAPA-TDNN}\label{ssec:ecapa}
Desplanques et al. \cite{Desplanques2020ecapa} propose several enhancements to the x-vector architecture. 
Their ECAPA-TDNN system adds 1-dimensional Res2Net \citep{gao2021res2net} modules with skip connections as well as squeeze-excitation (SE) blocks \citep{jie2018squeeze} to capture channel interdependencies. 
Furthermore, features are aggregated and propagated across multiple layers. 

The architecture also introduces a channel-dependent self-attention mechanism that uses a global context in the frame-level layers and the statistics pooling layer. 
It captures the importance of each frame given the channel and is used to compute a weighted mean and standard deviation for the channel. 
The final output of the pooling layer is a concatenation of the channel-wise weighted mean and standard deviation vectors.

We use the ECAPA-TDNN implementation from the SpeechBrain toolkit \citep{speechbrain}.  
The model receives 80-dimensional MFCCs as its input. 
The data processing pipeline is similar to the one described in \Cref{sec:xvec}. 
The system was trained using the VoxCeleb dataset, which was augmented by adding noise and reverberation. 
Additionally, the data is speed-perturbed at 95\% and 105\% of the normal utterance speed, and the SpecAugment \citep{specaugment2019} algorithm is applied in the time domain. 

\subsubsection{wav2vec 2.0}\label{ssec:w2v}
\begin{figure*}[!htb]
    \centering
    \includegraphics[width=\linewidth]{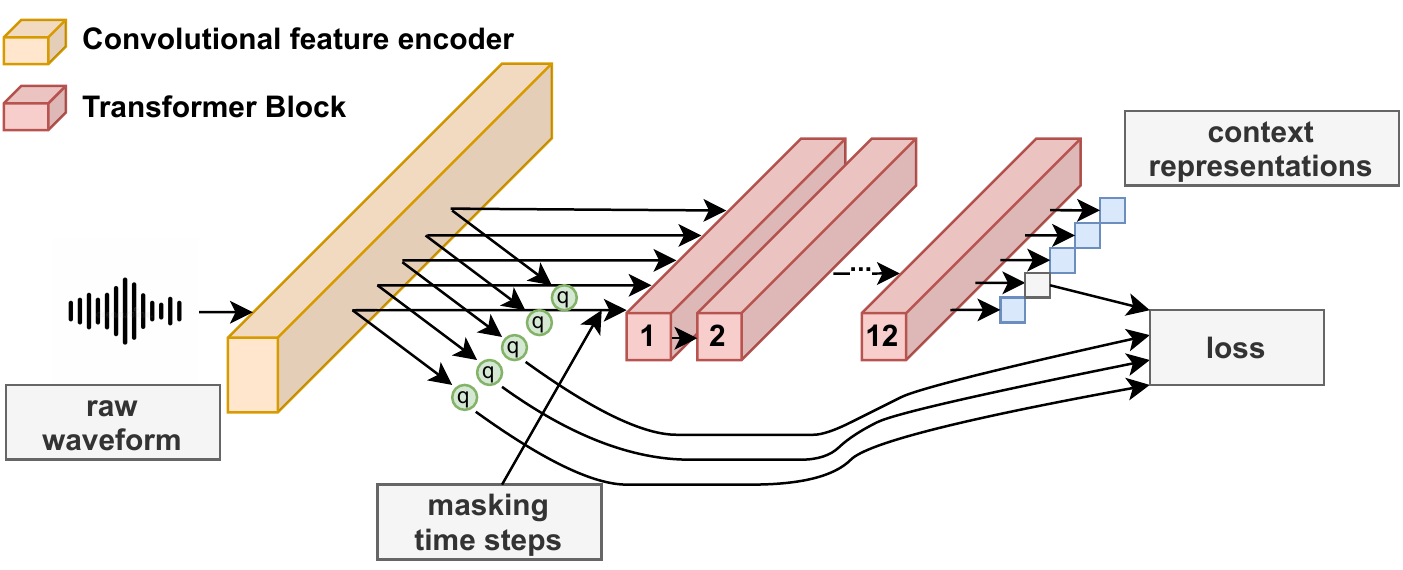}
    \caption{Schematic overview of the wav2vec 2.0 pre-training process based on \citep{baevski_wav2vec_2020}.
    The raw waveform is processed by a CNN that yields latent representations that are quantized (during pre-training only).
    The quantized representations are used for loss computation. 
    The latent features are further processed by the transformer encoder layers, but a portion of the time steps is masked. 
    The networks' objective is to predict the features at the masked time steps. 
    The contrastive loss is calculated using the output of the network (contextual representations) and the quantized representations.
    Once the model is pre-trained, it takes a raw waveform as input and yields hidden representations after each transformer block. 
    }
  \label{fig:w2v2_architecture}
\end{figure*}
Wav2vec 2.0 (W2V2) is a model based on the transformer architecture \citep{vaswani_attention_2017}.
Transformer models make heavy use of self-attention blocks, which help the model to focus on the ``most important'' parts of the input signal to represent the speech audio \citep{vaswani_attention_2017,baevski_wav2vec_2020}.
W2V2 was designed to learn speech features from large amounts of unlabeled training data.

The system is pre-trained in an unsupervised manner, i.e, unlabeled speech data is used for training. 
During pre-training, the objective is to predict masked quantized samples from a given signal context as visualized in \Cref{fig:w2v2_architecture}.
The system takes the raw audio waveform as its input, which is then processed by a convolutional neural network (CNN) feature encoder, yielding latent representations. 
The latent representations are then quantized and a portion of the time steps are masked simlarly to the pre-training of language models \citep{devlin_bert_2019}. 
The latent representations with the masked time steps are then processed by $n$ transformer encoder blocks. 
The last encoder layer then yields contextualized speech representations, which are then used to compute the loss w.r.t. the quantized representations.
Pre-trained models are commonly shared amongst researchers as the process of pre-training is computationally very expensive. 

The pre-trained W2V2 transformer models are widely used as general-purpose feature extractors for speech tasks and became the de-facto standard input to acoustic models for automatic speech recognition (ASR). 
In our experiments, we use a model pre-trained on 960 hours of unlabeled speech from the LibriSpeech corpus \citep{panayotov_librispeech_2015}, which was fine-tuned for ASR on the transcripts of the same data. 
The LibriSpeech corpus contains utterances from 2,484 different speakers. 

The W2V2 model yields intermediate representations after each of its 12 transformer blocks. 
A 768-dimensional vector is provided for approximately every 20ms of input audio. 
We extract those vectors for each 3-second chunk of audio and compute the mean over time for each of the 768 dimensions, yielding one vector representing three seconds
of audio. 
Intermediate representations extracted at different layers of a pre-trained model have been found to be suitable for varying tasks \citep{baevski_unsupervised_2021}.

\subsection{openSMILE}\label{sec:opensmile}
OpenSMILE is a toolkit that is used for the extraction of audio features \citep{eyben_OpensmileMunichVersatile_2010}.
The toolkit computes a set of speech features that represent the audio from which it is extracted, either on a recording level or over an audio segment of window length $l$. 
There are several pre-configured feature sets provided by the toolkit. 
The feature sets consist of statistical functionals (e.g., mean, median, and standard deviation) over low-level descriptors (LLDs) that describe the dynamic properties of the underlying signal contours. 

In our experiments, we use the ComParE 2016 feature set, which consists of 6373 static features representing the audio inside the extraction window.
OpenSMILE features are particularly useful for baseline experiments, as they describe the signal using interpretable and speech-related features.
The feature set has been shown to achieve adequate baseline performance in several paralinguistic applications such as gender detection, age detection, or speech emotion recognition  \citep{schuller_interspeech_2016,schuller_ACMMultimedia2022_2022}. 

\subsection{Temporal Smoothing and Recording Normalization}\label{ssec:smooth}
We propose temporal smoothing and recording-level embedding normalization as pre-processing steps. 
Temporal smoothing averages embeddings along the time dimension using a sliding window with overlap. 
We define a window of length $w$ that is used to select consecutive vector representations $\mathbf{v}$ starting at time $i$. 
The arithmetic mean along each channel is computed for all $\mathbf{v}$ under the current window. 

Given a sequence of $W$ embeddings $\lbrace \mathbf{v}_{i}\rbrace_{i=1}^W$, the smoothing operation yields a new sequence $\lbrace \mathbf{s}_i \rbrace_{i=1}^{W-w+1}$ by computing the mean over subsequences of $w$ terms:
\begin{equation}
     \mathbf{s}_i=\frac{1}{w} \sum_{j=i}^{i+w-1} \mathbf{v}_j. 
\end{equation}

This procedure effectively masks characteristics in the latent representations that are specific to a point in time and allows our classifiers to focus more on changes that occur gradually over time. 
We use window lengths of 30 and 60 seconds in our experiments. 
Since each embedding represents 3 seconds of audio, the window length $w$ measured as the number of elements considered for averaging is given by $w \in \{30/3, 60/3\}$. 
To match this experimental setup, we also perform temporal smoothing on openSMILE features using the same parameters based on openSMILE features with an extraction context of 3s.
Due to the nature of the openSMILE features,
being a statistical description of the contours of the underlying LLDs, we also experimented with changing the extraction window $w \in \{3, 6, 30, 60, 120, 180\}$.
Longer extraction windows do not make sense in this scenario, as it would yield too few training samples. 
The features are extracted differently than the neural embeddings and do not model relationships between neighboring segments.

The second pre-processing step can be described as a type of recording-level normalization. 
The ECAPA-TDNN and x-vector models employed here were designed for speaker identification tasks. 
Hence, latent representations extracted from hidden layers of these systems can be expected to primarily encode speaker-specific traits. 
However, we are interested in subtle phonological differences encoded in embeddings extracted at different points in time. 
Therefore, we compose prototypical x-vectors and ECAPA-TDNN embeddings over an entire lecture recording, i.e, we receive a single vector for each lecture that encodes aggregated speaker-specific characteristics. 
These prototypical representations are then subtracted from each embedding.
The primary goal of this approach is to enhance characteristics related to the speaker's voice and to filter out traits that are shared across the entire recording. 
For a sequence of $W$ embeddings, each representing a short chunk of the lecture $\lbrace \mathbf{v}_{i}\rbrace_{i=1}^W$, we obtain a new sequence of normalized embeddings $\lbrace \mathbf{n}_i \rbrace_{i=1}^{W}$ by subtracting a constant prototype $\mathbf{p}$:
\begin{equation}\label{eq:proto}
     \mathbf{n}_i= \mathbf{v}_i - \mathbf{p}. 
\end{equation}
For the x-vector and ECAPA-TDNN systems, $\mathbf{p}$ is a representation generated by passing spectral features of the entire recording to the model. 
The prototype vector for a complete lecture recording using W2V2 could not be directly computed due to memory limitations. 
In this case, $\mathbf{p}$ is constructed by computing the arithmetic mean over all extracted embeddings per lecture, which is then subtracted from all other W2V2 vectors. 
The prototypical vector for the openSMILE features is also computed as the arithmetic mean of all extracted vectors, as the features included lose their meaningfulness with longer extraction windows. 

\section{Experiments}\label{sec:experiments}
\begin{figure}[htb]
    \centering
    \includegraphics[width=0.95\linewidth]{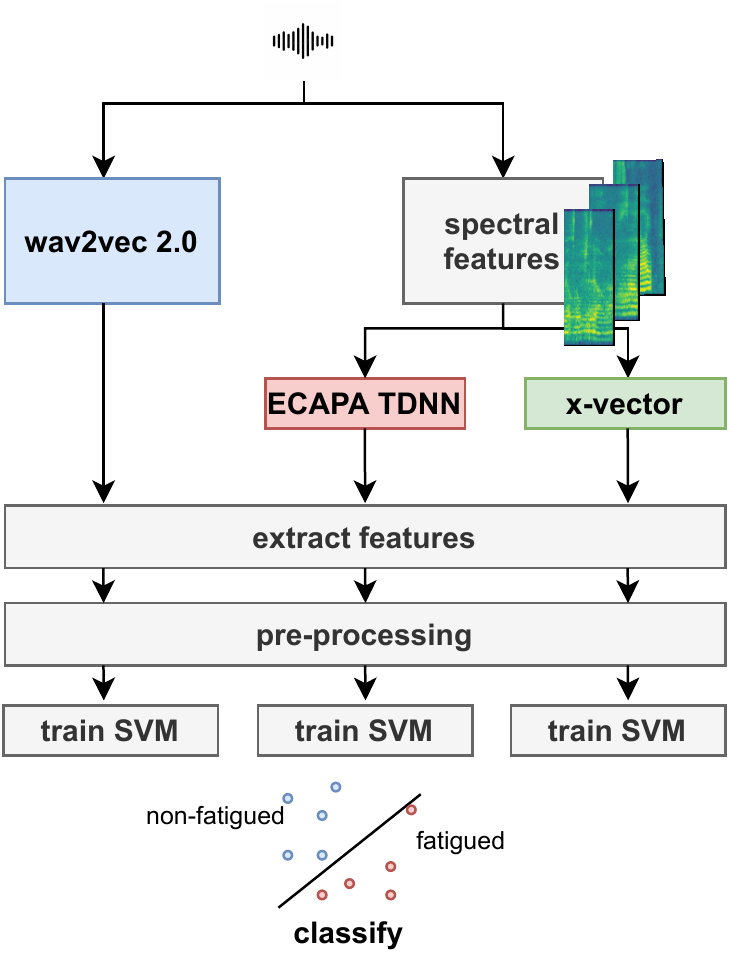}
    \caption{Schematic overview of the training and classification pipeline starting from the raw waveform. The preprocessing step contains recording-level normalisation, temporal smoothing and mean pooling of embeddings of the outputs of the pre-trained wav2vec 2.0 features over 3-second snippets.}
  \label{fig:flow_diag}
\end{figure}

A schematic overview of the processing and classification pipeline can be found in \Cref{fig:flow_diag}.
We divided each lecture into non-overlapping 3-second segments of audio, which were then passed to the respective model for embedding retrieval. 
The x-vector and ECAPA-TDNN models yield a single embedding for each audio segment that is passed to the extractor. 

SVMs are effective on high-dimensional data, they tend to perform well on datasets of small to moderate size, and in comparison to methods with many tunable free variables (e.g. neural networks), they are less prone to overfitting. 
Furthermore, SVMs have already been applied to vocal fatigue detection \citep{CARATY2014453}, and are commonly used in conjunction with neural embeddings for other pathological voice classification tasks \citep{braun22_theft,bayerl_DetectingDysfluenciesStuttering_2022a,botelho22_interspeech}. 
Therefore, we chose SVMs for the downstream classification task.


\subsection{Visualization}
\begin{figure*}[htb]
    \centering
    \includegraphics[width=0.95\linewidth]{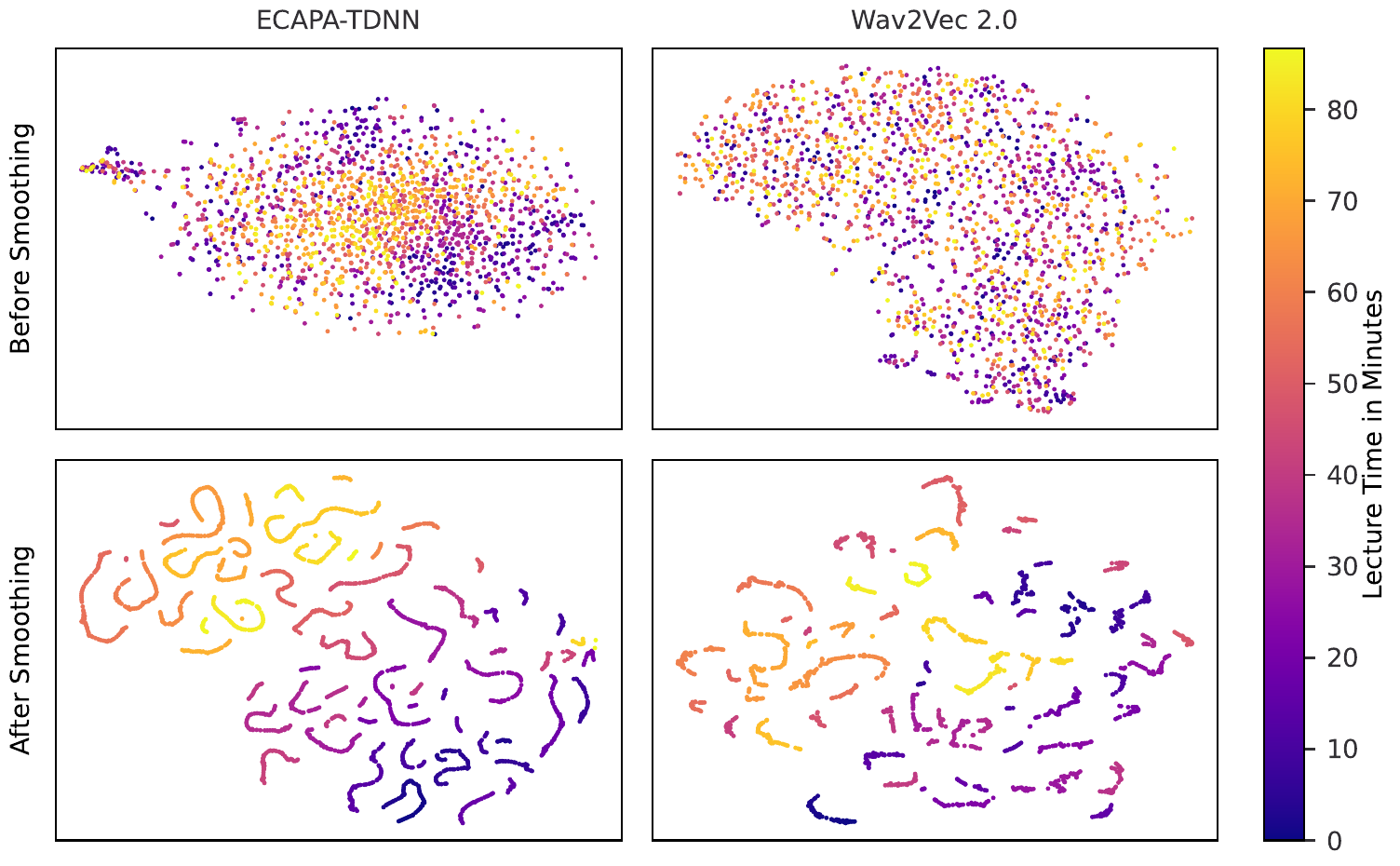}
    \caption{t-SNE transform of ECAPA-TDNN and W2V2 embeddings extracted from lecture PA13 of the LMELectures corpus before and after smoothing with a window length of 30 seconds. The W2V2 embeddings have been extracted at the first layer. The ECAPA embeddings already show some structure before applying temporal smoothing but still show a lot of overlap in the visualization. After temporal smoothing both feature types show similar local worm-like structures and also show that there is a greater global structure, showing differences w.r.t. the time in the recording they were extracted from.} 
  \label{fig:ecapa_comparison}
\end{figure*}

Using dimensionality reduction techniques to visualize low-dimensional projections of high-dimensional data can lead to important insights about the structure of the underlying data. 
We mapped the high-dimensional model outputs (x-vector $\mathbb{R}^{512}$, ECAPA-TDNN $\mathbb{R}^{192}$, W2V2 $\mathbb{R}^{768}$) to locations in two-dimensional space by means of a t-SNE transform. 
An example of these two-dimensional representations is depicted for lecture \textit{PA13} in \Cref{fig:ecapa_comparison}.
The first row of \Cref{fig:ecapa_comparison} shows how ECAPA-TDNN (left column) and W2V2 (right column) embeddings are distributed after applying t-SNE with a perplexity parameter of $ppl=30$. 
Each dot represents the transformed version of an embedding at a certain time. 
The colors indicate the lecture's progress. 
Blue dots represent the first half of the lecture (up to 40 minutes), while red dots represent the second half of the lecture. 
The distribution of the dots shows that embeddings that are close to each other in the time domain, are also packed together in the figure, i.e., blue dots can be found together with other blue dots, and red dots accompany other red dots. 
The two-dimensional representations of ECAPA-TDNN embeddings (upper left) show more pronounced temporal clusters than the W2V2 representations (upper right). 

The effect of the temporal smoothing method described in \Cref{ssec:smooth} is illustrated in the second row of \Cref{fig:ecapa_comparison}. 
In this case, the channel-wise mean over all embeddings in a 30-second sliding window is computed. 
Hence, ten distinct embeddings, each representing 30 consecutive seconds of lecture audio at a resolution of 3 seconds, are aggregated to a single high-dimensional representation and are then mapped into two-dimensional space via t-SNE.  
\Cref{fig:ecapa_comparison} shows that the various stages of the lecture (e.g. first part, middle part, last part) are more distinguishable when smoothing is applied (second row), compared to no pre-processing (first row). 
Temporal smoothing leads to worm-like local structures for different periods in the lecture (e.g. the first 10 minutes indicated by dark blue colors). 
Those local clusters are embedded in a wider global structure roughly dividing the lecture in its first and second half. 
\subsection{Classification}
\setlength{\tabcolsep}{5pt} 
\renewcommand{\arraystretch}{1} 
\begin{table}[th]
  \caption{Results of binary classification experiments. 
  We report the best results on the test set comprised of \textit{PA} lectures with model parameters obtained via grid search w.r.t accuracy. 
  The columns \textit{NF} and \textit{F} represent non-fatigued and fatigued speech, respectively. 
  The column \textit{Win.} refers to the length of the temporal smoothing window in seconds for neural embeddings (Xvec, ECAPA, W2V2) and openSMILE with temporal smoothing (oS-TS) or the extraction window for openSMILE features.}
  \label{tab:results}
  \centering
  \begin{tabular}{lllccccc}
    \toprule
	\# &  Emb. &      Win.  &  \multicolumn{2}{c}{Precision} & \multicolumn{2}{c}{Recall} & Acc. \\
     &                       &         (sec.)    &  NF & F  & NF & F  & \\
    \midrule
	\multicolumn{8}{c}{\textbf{No Normalization}}\\
	\hline
    \textit{1} &\multirow{5}{*}{openSMILE} & 3  & 0.53 & 0.52  & 0.44 & 0.60   & 0.53 \\
    \textit{2} &                      & 6       & 0.58 & 0.64  & 0.72 & 0.49   & 0.60 \\
    \textit{3} &                      & 30       & 0.57 & 0.95  & 0.99 & 0.24   & 0.61 \\
    \textit{4} &					  & 60       & 0.58 & 0.58  & 0.56 & 0.61   & 0.58 \\
    \textit{5} &					  & 120       & 0.62 & 0.65  & 0.70 & 0.56   & 0.63\\
    \textit{6} &					  & 180       & 0.53 & 0.55  & 0.63 & 0.44   & 0.54\\
	\hline
   \textit{7} & \multirow{2}{*}{oS-TS} &  30 &  0.59 &         0.52 &      0.21 &      0.85 &      0.53 \\
   \textit{8} &				  & 60       &   0.60 &         0.54 &      0.37 &      0.75 &      0.56 \\
    
	\hline
    \textit{9} &\multirow{3}{*}{Xvec} & --       & 0.66 & 0.56  & 0.37 & 0.81   & 0.59\\
    \textit{10} &                      & 30       & 0.85 & 0.59  & 0.34 & 0.94   & 0.64\\
    \textit{11} &					  & 60       & 0.87 & 0.59  & 0.35 & 0.95   & 0.65\\
    \hline
    \textit{12} &\multirow{3}{*}{ECAPA}& --       & 0.73 & 0.61  & 0.46 & 0.83   & 0.65\\
    \textit{13} &                      & 30       & 0.85 & 0.63  & 0.47 & 0.62   & 0.69\\
    \textit{14} &					  & 60       & 0.83 & 0.66  & 0.54 & 0.89   & 0.72\\
    \hline
    \textit{15} & \multirow{3}{*}{W2V2} & --       & 0.73 & 0.60    & 0.44  & 0.84 & 0.64 \\
    \textit{16} &					  & 30       & 0.81 &	0.64  & 0.50 & 0.89 & 0.69 \\
    \textit{17} &					  & 60       & 0.81 &	0.63  & 0.47 & 0.89 & 0.68 \\
    \hline
	\multicolumn{8}{c}{\textbf{Recording Normalization}}\\
	\hline
    \textit{18} &\multirow{5}{*}{openSMILE} & 3 & 0.54 & 0.54  & 0.50 & 0.58   & 0.54 \\
    \textit{19} &                      & 6      & 0.63 & 0.61  & 0.59 & 0.65   & 0.62 \\
    \textit{20} &                      & 30      & 0.64 & 0.67  & 0.70 & 0.61   & 0.65 \\
    \textit{21} &					  & 60       & 0.64 & 0.63  & 0.62 & 0.64   & 0.63 \\
    \textit{22} &					  & 120       & 0.74 & 0.74  & 0.74 & 0.74  & \textbf{0.74} \\
    \textit{23} &					  & 180       & 0.54 & 0.55  & 0.59 & 0.50  & 0.55 \\
	\hline
    \textit{24} & \multirow{2}{*}{oS-TS}  & 30 & 0.52 &         0.52 &      0.55 &      0.48 &      0.52 \\
    \textit{25} &					      & 60               & 0.51 &         0.52 &      0.54 &      0.49 &      0.52 \\
    \hline
    \textit{26} &\multirow{3}{*}{Xvec}   & --       & 0.65 & 0.63   & 0.60 & 0.67   & 0.64\\
    \textit{27} &                        & 30       & 0.80 & 0.75   & 0.72 & 0.81   & 0.77\\
    \textit{28} &						 & 60       & 0.85 & 0.77   & 0.74 & 0.87   & \textbf{0.81}\\
    \hline
    \textit{29} &\multirow{3}{*}{ECAPA} & --        & 0.69 & 0.69   & 0.70 & 0.68   & 0.69\\
    \textit{30} &                       & 30        & 0.78 & 0.84   & 0.85 & 0.76   & 0.81\\
    \textit{31} &					    & 60        & 0.82 & 0.90   & 0.91 & 0.80   & \textbf{0.85} \\
    \hline
    \textit{32} & \multirow{3}{*}{W2V2} & --      & 0.67 & 0.68      & 0.69 & 0.66 &  0.68 \\ 
    \textit{33} &					    & 30       & 0.80 & 0.81      & 0.81 & 0.80 & 0.80 \\
    \textit{34} &					    & 60       & 0.84 & 0.81      & 0.80 & 0.85 & \textbf{0.82} \\
    \bottomrule
  \end{tabular}
\end{table}
Following Caraty and Montacié \cite{CARATY2014453}, we define a binary classification task, in which the first segment of a lecture with duration $d$ is representative of the class ``non-fatigue'' (NF) and a later segment with the same duration is representative for the class ``fatigue'' (F). 
We set $d=10$ minutes and assign all embeddings from minute 0 to minute 10 to class \textit{NF} and all embeddings starting at minute 50 and ending at minute 60 to class \textit{F}.   
SVMs were trained using radial basis function (RBF) kernels. 
The optimal hyperparameters for the estimator were determined with the grid search method in a fivefold cross-validation on the training set.
Principal component analysis (PCA) is performed prior to SVM-training to reduce dimensionality. 
The number of principal components is chosen from $N_{pca} \in \{ 2^{k} \mid k = 5, \ldots, \lfloor \log_2 D\rfloor \} \subset \mathbb{N}_{>0}$, where $D$ is the dimensionality of the embeddings. 
The kernel parameter $\gamma$ is selected from the set $\gamma \in \{10^{-k} \mid k = 5, \ldots, 1 \} \subset \mathbb{R}_{>0}$, and
the penalty parameter of the error term $C$ is selected from $C \in \{ 5, 10, 20, 50 \} \subset \mathbb{N}_{>0}$.
Hyperparameter optimization and training were performed on the \textit{IMIP} lectures, while the \textit{PA} lectures were used for testing. 
We conducted multiple classification experiments with and without recording normalization as well as varying smoothing window lengths ranging from 0 (no smoothing) up to 60 seconds. 
Since we were interested in capturing gradual changes over time, the window lengths were chosen to smooth potential variability in the articulation rate, which can be substantial in spontaneous speech \citep{Miller1984}, while still allowing for variation over longer periods of time. 
Furthermore, we expect that these window lengths ensure sufficient phonetic coverage.

\subsection{Results}
The baseline results with openSMILE features in \Cref{tab:results} show that using an extraction window of at least 6 seconds led to accuracy rates of more than 50\% in the vocal fatigue classification task. 
Applying temporal smoothing in the same manner as to the neural embeddings did not improve results using openSMILE.
The best accuracy (74\%) was achieved with an extraction window of 120 seconds and recording-level normalization. 
A previous study using the ComParE 2010 feature set (1582 features) to detect the change of voice characteristics during prolonged voice utilization only reported above chance level results with extraction window sizes of greater than 120 seconds when evaluating on the test set \citep{CARATY2014453}.

The classifier trained on ECAPA-TDNN embeddings with a smoothing window length of 60 seconds yielded the best overall accuracy of 85\% with recording normalization (cf. experiment \#27 in \Cref{tab:results}).
Using temporal smoothing and recording normalization led to performance improvements for all three types of embeddings. 
For example, recording normalization improved the accuracy scores of the best models by 25\% (x-vector), 18\% (ECAPA-TDNN), and 21\% (W2V2). 
Accuracy generally increased with increasing smoothing window length. 
For example, temporal smoothing applied on x-vector embeddings with recording normalization led to a relative improvement of 27\%. 
The effect on ECAPA-TDNN and W2V2 embeddings was slightly less pronounced with relative improvements of 23\% and 21\%. 
However, we also noticed that window lengths of more than 60 seconds started to have a negative effect on classification performance. 

We applied SVMs trained on all LMELectures (\textit{IMIP} and \textit{PA}) with a duration of more than 60 minutes, using the set of hyperparameters that led to the best results in \Cref{tab:results}, on the additional DL lectures corpus introduced in \Cref{sec:data}. 
The classifiers yielded accuracy scores of 72\% (x-vector), 70\% (ECAPA-TDNN), and 76\% (W2V2). 

\begin{figure}[!htb]
    \centering
    \includegraphics[width=\linewidth]{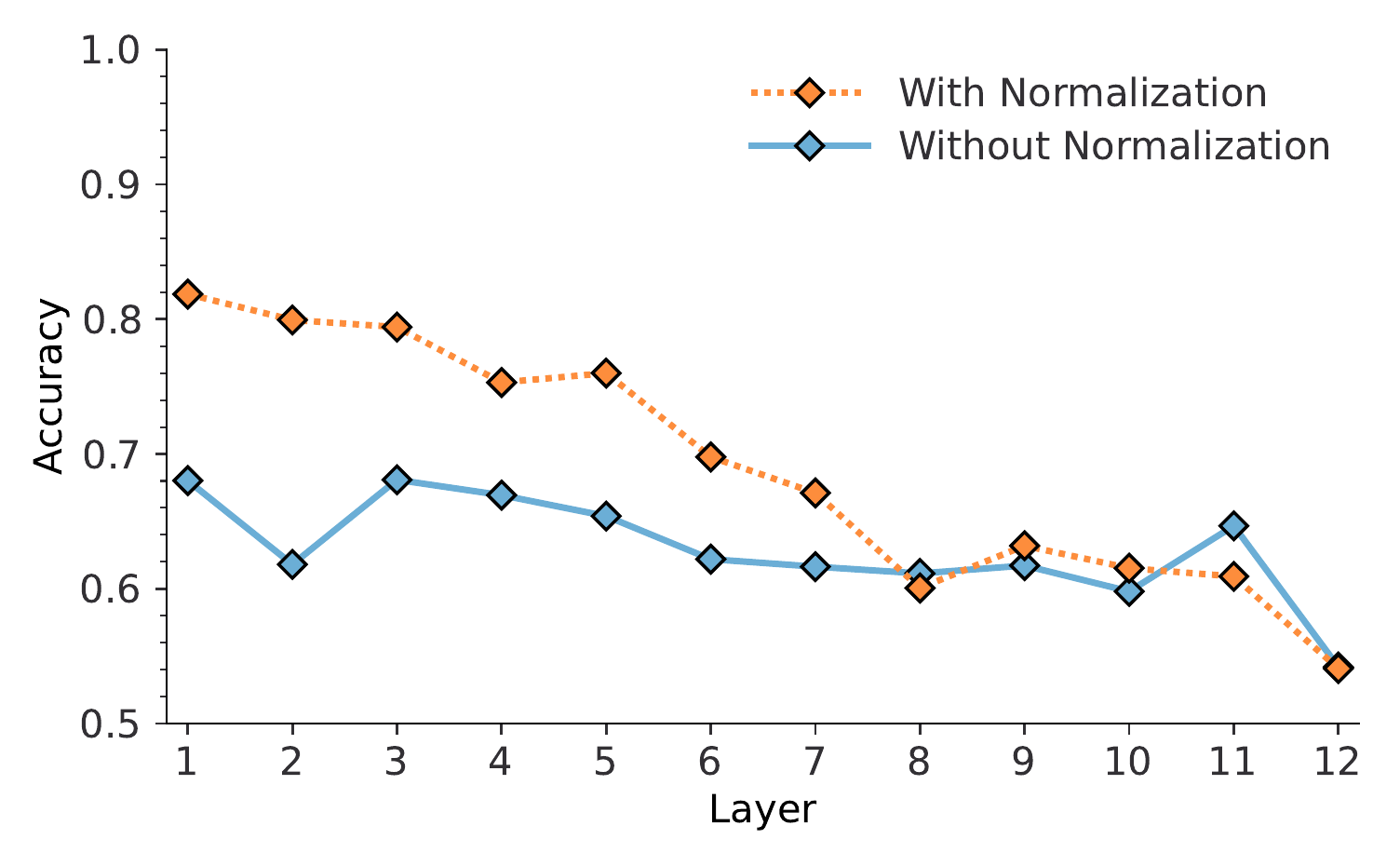}
    \caption{Classification performance on \textit{PA} lectures with embeddings from different W2V2 layers.}
  \label{fig:w2v2performance}
\end{figure}

\Cref{fig:w2v2performance} shows the layerwise performance of systems trained on W2V2 features with and without recording-level normalization. 
\subsection{Discussion}

The classification results presented in \Cref{tab:results} indicate that all three types of neural embeddings can be used as features to detect vocal fatigue. 
All types of embeddings outperform the baseline method based on openSMILE features substantially and operate on shorter time-frames, making them more practically applicable. 

The experiments on the additional \textit{DL} lectures corpus show that the models generalize well to a similar speaker and similar recording conditions. 
However, due to these similarities, statements about the overall robustness of the systems are still limited. 
Based on the available data, we cannot make statements about the generalizability of our models to a larger number of speakers across different age cohorts and genders. 
Furthermore, this study relies on conditions that might not always be given practice, such as high-quality recordings and constant recording environments. 

The system trained on W2V2 embeddings outperformed the systems trained on the x-vector and ECAPA-TDNN embeddings, when confronted with an unseen speaker. 
An explanation for the better performance might be that W2V2 was primarily designed for ASR tasks, which aim to work independently of the speaker and therefore attenuate speaker characteristics.
On the other hand, ECAPA-TDNN and x-vector embeddings were specifically designed to emphasize speaker characteristics. 
We hypothesize that the systems trained on x-vector and ECAPA-TDNN representations capture features that are more relevant to the speaker in the training set, whereas the systems based on W2V2 embeddings are capable of learning more general characteristics related to the change of voice after prolonged speaking. 

As stated by Baevski et al. \citep{baevski_unsupervised_2021} and others, embeddings extracted at different layers of W2V2 are suitable for different tasks.
Without recording-level normalization, this was barely reflected in the accuracy scores depicted in \Cref{fig:w2v2performance} (solid blue line).
However, once normalization was applied, the differences became more pronounced.
Layers 1-4 performed well, with a steep decline in performance after layer 5 (dotted orange line).

Performance fatigue cannot be interpreted as a singular event or moment from which a person's voice is fatigued. 
There should be a gradual process from non-fatigued to fatigued that can be detected because of changes in voice characteristics over time due to prolonged utilization.
Technical means to detect and measure vocal fatigue can only pick up those changes once these are prominent enough.  
For practical applications, this would mean figuring out when changes in voice characteristics are detectable. 
The experiments in this paper, differentiating only between fatigued and non-fatigued, represent a simplified view of this problem, assuming the changes over the course of 40 minutes are detectable. 
The question of when the change in voice characteristics is prominent enough to be detected remains unanswered and can be assumed to be speaker dependent. 

\begin{figure}[ht]
   \includegraphics[width=\linewidth,right]{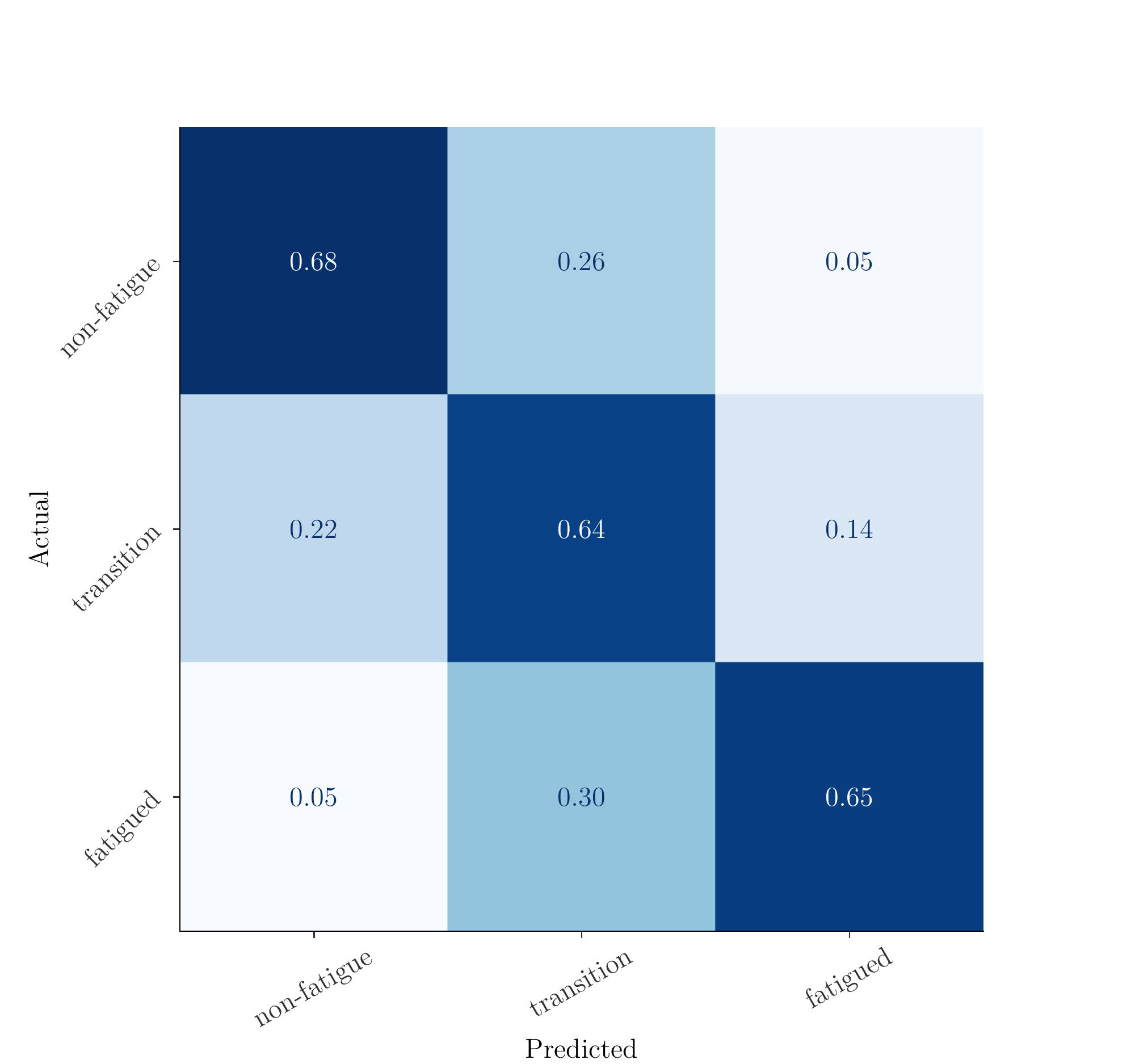}
    \caption{Confusion matrix for a three-class classification experiment using SVMs and wav2vec 2.0 embeddings with recording-level normalization and temporal smoothing.}
    \label{fig:conf_mat_three}
\end{figure}

\Cref{fig:conf_mat_three} shows a confusion matrix for a three-class classification problem, using embeddings taken from the first ten minutes, the ten minutes in the middle of the lecture, and the last ten minutes. 
The confusion matrix clearly shows that almost no misclassifications are happening from fatigue to non-fatigued and vice versa, supporting this assumption.
The transition class gets confused in both directions, slightly favoring the non-fatigued class. 
This supports the hypothesis of a gradual change and suggests further experiments using regression rather than classification. 

This study addresses vocal fatigue from an objective perspective based on voice characteristics embedded in neural representations. 
We rely on the assumption that the characteristics of vocal fatigue are encoded in high-dimensional neural embeddings and that these characteristics change after prolonged periods of 
 voice utilization.

The underlying feature extractors are trained to capture speaker-specific information and suppress other factors, such as recording conditions. 
Furthermore, our feature extractors have seen many speakers (at least 2k speakers) during training. 
Therefore, a certain robustness against speaker variation can be assumed, but the exact influence of speaker variations on classification performance remains to be explored. 
\section{Conclusions and Future Work}
We demonstrated that x-vectors, ECAPA-TDNN embeddings, and W2V2 embeddings can be used to reliably classify
speech into  ``fatigue'' and ``non-fatigue''.
The classifier trained on ECAPA-TDNN embeddings with normalization and temporal smoothing (window length of 60 seconds) yielded the best overall accuracy of 85\% on the test set. 
The results also showed that temporal smoothing and recording normalization improved overall performance.
Our classifiers generalized to an unseen speaker and recording environment without adaptation, achieving accuracies between 70\% and 76\% on the DL lectures corpus. 
Our approach is limited by the features that are encoded in neural embeddings. 
However, the results of empirical studies indicate that psychological and environmental factors also play a role in the occurrence of vocal fatigue \citep{Gotaas1993teacher,Cercal2020}. 
These studies also show that the severity of perceived vocal fatigue of teachers is significantly higher at the end of a workday and that university professors perceive a higher degree of vocal fatigue at the end of a term. 
Therefore, we cannot make statements about the extent to which other drivers of vocal fatigue influenced the results presented here.
As such the methods described in this paper detect only one aspect of a complex phenomenon but do so reliably. 

The focus during the creation of the training and test corpora used in this study was the minimization of factors that might distort the classification results, such as varying recording conditions or changes in speaking style. 
However, this introduced several other limitations (cf. \Cref{ssec:limitations}) that prohibit conclusions about the overall generalizability of our approach. 
Therefore, next steps will include the creation of an expanded corpus that includes male and female speakers across a wide age range, as well as subjective fatigue labels. 
Future work will then explore the correlation between the approach presented here and subjective labels. 
In addition, this enables a comprehensive analysis of the robustness across a wide range of speakers and recording environments. 
Additionally, we will strive towards more granular predictions (e.g., multiclass classification or regression) and overall performance improvement by taking the above factors into account. 
Further research will also compare the performance differences between a single speaker-independent model and multiple speaker-dependent models. 

\section{Acknowledgements}
The authors would like to express their gratitude to Andreas Maier for kindly giving permission to include his lecture recordings in this study. 
This work was partially funded by the Bavarian State Ministry of Science and supported by the Bayerisches Wissenschaftsforum (BayWISS).

\FloatBarrier
\bibliographystyle{elsarticle-num} 
\bibliography{references,refs,zotero}
\end{document}